\documentclass[10pt, conference, letterpaper]{IEEEtran}

\newif\ifisanon
\isanonfalse

\IEEEoverridecommandlockouts
\usepackage{cite}
\usepackage{tikz}
\usepackage{amsmath}

\usepackage{graphbox} 

\usepackage{filecontents}

\usepackage{url}
\usepackage{multicol}
\usepackage{subcaption}
\usepackage{graphicx} 
\usepackage{algorithmicx} 
\usepackage{algorithm2e}
\usepackage{hyperref}

\usepackage{tikz} 
\usepackage{pgf}
\usetikzlibrary{arrows,automata}
\usepackage[all]{xy}
\usepackage{pbox} 
\usepackage{comment}
\usepackage{amsfonts}
\usepackage{adjustbox}
\hypersetup{colorlinks,citecolor=blue,bookmarks=false,breaklinks}

\def\Snospace~{\S{}}
\newcommand\V[1]{{\mbox{\textit{#1}}}}

\makeatletter
  \renewcommand\comment[1]{{\color{blue} \sffamily [xxx:  #1]}}

  \newcommand\reviewfix[1]{{\sffamily [RF:#1]}\@bsphack\@esphack}
  \newcommand\PostSubmission[1]{{\sffamily [post-xxx:  #1]}}
  \iftrue
    \renewcommand\comment[1]{}
    \renewcommand\PostSubmission[1]{}
    \renewcommand\reviewfix[1]{\@bsphack\@esphack}
  \fi
\makeatother

\def\BibTeX{{\rm B\kern-.05em{\sc i\kern-.025em b}\kern-.08em
    T\kern-.1667em\lower.7ex\hbox{E}\kern-.125emX}}
\begin{document}


\title{Third-Party Assessment of Mobile Performance \\in the 5G Era}

\author{\IEEEauthorblockN{ASM Rizvi$^{*}$}

\IEEEauthorblockA{
\textit{Akamai Technologies}\\
asrizv@akamai.com}

\and
\IEEEauthorblockN{John Heidemann}
\IEEEauthorblockA{
\textit{USC/ISI}\\
johnh@isi.edu}
\and
\IEEEauthorblockN{David Plonka$\dagger$}
\IEEEauthorblockA{
\textit{WiscNet}\\
plonka@wiscnet.net}

}

\maketitle
\renewcommand{\thefootnote}{\fnsymbol{footnote}}
\footnotetext[1]{Part of this work was conducted while ASM Rizvi was a PhD student at USC/ISI.}
\footnotetext[2]{This research was carried out when David Plonka was affiliated with Akamai Technologies.}

\begin{abstract}
The web experience using mobile devices
  is important since a significant portion of the
  Internet traffic is initiated from mobile devices.
In the era of 5G, users expect a 
  high-performance data network to stream media content and
  for other latency-sensitive applications.
In this paper,
  we characterize mobile experience in terms of
  latency, throughput, and stability measured
  from a commercial, globally-distributed CDN. 
Unlike prior work, 
  CDN data provides a relatively neutral, carrier-agnostic perspective, 
  providing a clear view of  multiple and international providers.
Our analysis of mobile client traffic shows 
  mobile users sometimes experience markedly low latency, 
  even as low as 6\,ms. 
However, only the top 5\% users 
  regularly experience less than 20\,ms of minimum latency.
While 100\,Mb/s throughput is not rare,
  we show around 60\% users observe less
  than 50\,Mb/s throughput.
We find the minimum mobile latency is generally stable
  at a specific location which can be an important characteristic 
  for anomaly detection.
\end{abstract}


\section{Introduction}

Mobile providers today offer increasingly 
  high-speed Internet service~\cite{cellular-growth, cellular-forecast}.
They aim to provide
  low latency and high throughput 
  to support
  multimedia streaming, Internet-of-Things (IoT) connectivity, and vehicle-to-vehicle (V2V) communication.
To fulfill these service requirements,
  they have added new technologies in
  radio spectrum (mmWave), edge computing, and network slicing.
Today's 5G theoretically provides up to 20\,Gb/s throughput~\cite{5g-performance, an2023adaptive, pogge2019enabling}
  and end-to-end latencies as targeting 2\,ms~\cite{5g-lat-thr, jun2020ultra}.
\comment{I dropped ``in the mobile domain'' because ``end-to-end'' is not---johnh 2024-03-14}
However, achieving the theoretical best in practice remains elusive. 
\comment{I don't think we can say 2ms and then 10ms here... we're fighting with ourselves.  I removed the next sen ---johnh 2024-03-14}
\comment{was: ``Next generation networks target end-to-end latency below 10\,ms~\cite{5g-expect}.'' ---johnh 2024-03-14}

While 5G allows new capabilities,
  how quickly do 5G operators deploy them,
  and how available are they to users?
Market pressures encourage rapid deployment of ``5G'', 
  but early hardware may not include all features,
  and operators may delay feature availability 
  while they gain confidence in their stability.
New features often must be explicitly enabled,
  and operators may delay feature roll-out pending
  integration with new billing models
  or specific commercial opportunities.

After several years of global 5G deployment, 
  our goal is to assess the actual performance of 5G networks, 
  both to gauge their current status and 
  to explore their potential.

Content Delivery Networks (CDNs) provide a unique
  opportunity to provide a third-party assessment for 5G across
  multiple mobile operators.
CDNs are responsible for delivering popular content to users
  from their distributed infrastructure.
A globally distributed CDN receives traffic from almost
  all the mobile carriers around the globe. 
As a result, CDNs can observe the performance of the mobile users
  as a third-party observer, without requiring any direct measurement 
  from the mobile users.
Although direct measurements of specific CDN devices are valuable,
  broad measurements of many 5G users from a CDN
  can avoid potential bias that can arise from direct measurements
  of a few users.

In this paper,
  we characterize mobile latency and throughput and
  make two contributions.
Our first contribution is to describe an approach to identify existing mobile user equipment (UE) traffic measurements in a globally distributed CDN.
As CDN logs aggregate traffic from various devices, we rely on the IPv6 address pattern to differentiate mobile User Equipment (UE) from other sources (\autoref{sec:ipv6-to-get-cellular}).
Our second contribution is to
  characterize end-to-end mobile latency and throughput along
  with their stability (\autoref{sec:stability-mobile-network}).
We evaluate the limits of latency (\autoref{sec:latency}),
  throughput (\autoref{sec:throughput}), 
  and stability (\autoref{sec:stability})
  that clients can achieve.
Our goal is to examine the extent to which 5G approaches the 
  targets of throughput and latency, 
  as well as how closely the current latency aligns with 
  the anticipated expectations of 6G.

\textbf{Anonymization, Data, and Ethics:} 
We do not reveal the carrier names
  when we report the results from the CDN data,
  since our goal is to evaluate latency and throughput
  relevant to 5G targets, not between carriers.
Because CDN data is anonymized and reflects proprietary details,
  we regret that we cannot make our data available.
Our work poses no ethical concerns as described in \autoref{sec:ethics}.

\section{Related Work}
  \label{related-work}

Related studies explore measurements 
  from mobile UE and CDNs, and CDN performance.

\textbf{Measurement from UE:}
Previous studies showed
  performance measurement from 
  real mobile devices
  to evaluate 5G latency and throughput\cite{narayanan2020first, narayanan2021variegated, yuan2022understanding, ghoshal2023performance}.
Several studies took the mobile device
  to different locations in the US, and 
  measured latency and throughput while they moved~\cite{ghoshal2023performance, zhang2021inferring}.
Other studies
  measured latency and throughput
  within a limited geo-coverage from 
  the UE.
Some of these studies also measured
  latency, throughput, and power efficiency 
  with Stand-Alone (SA) and Non-Stand-Alone (NSA)
  5G networks~\cite{yuan2022understanding}.
\comment{``broader coverage'': please clarify: broader in what dimension? (what metric, how much?) ---johnh 2024-03-07}
\comment{modified. ---asmrizvi 2024-03-08}
By leveraging CDN logs instead, 
  we achieve broader coverage across a greater number of carriers and geolocations.

\textbf{Measurement from CDN:}
Closer to our work, one prior study compared SA-5G and NSA-5G
  delay, download speed, and energy consumption using a Chinese CDN with streaming capabilities~\cite{yuan2022understanding}.
We too study using a CDN,
  but we study global mobile phones and use a CDN with a global footprint.
While that work did a good job of evaluating CDN performance in China, 
  our work instead looks at the performance of mobile providers internationally,
  considering operators in four different countries from three continents.
\comment{ok, but is this a difference from that prior work?  as written it's not clear ---johnh 2024-03-07}
\comment{Tried one more time. ---asmrizvi 2024-03-08}
In contrast to the previous study, we utilize a globally distributed CDN. 
Considering the CDN's server deployment 
  near Mobile Edge Computing (MEC) facilities~\cite{hsu2020dns}, 
  the latency analysis will inform us about 
  the proximity of CDNs' server placement to the UE.

\textbf{CDN performance:}
Prior studies evaluated CDN performance to show that 
  the general Internet users of the CDN 
  see a good selection of CDN sites and client proximity
  to the nearby CDN front-end servers~\cite{calder2015analyzing, adhikari2014measurement, adhikari2012unreeling}.
One study confirmed most clients normally get
  their service from the nearby CDN site~\cite{calder2015analyzing}.
Another study showed how these CDNs are connected
  with a different number of peers~\cite{wohlfart2018leveraging}.
In this study,
  we show mobile latency and throughput 
  from a CDN perspective.
We demonstrate what UE can expect
  when they get their service from a global CDN.

\section{Architectural Considerations}
    \label{sec:cellular-variability}

\comment{We decided to keep our focus on end-to-end latency by the CDN. But this section had latency within cellular networks. I have addressed your comments and restructured this section a bit. ---asmrizvi 2024-03-08}
In this section, we describe the components of the mobile networks and 
  their interaction with edge computing and CDN servers.

\begin{figure}
\centering
\includegraphics[width=0.85\linewidth]{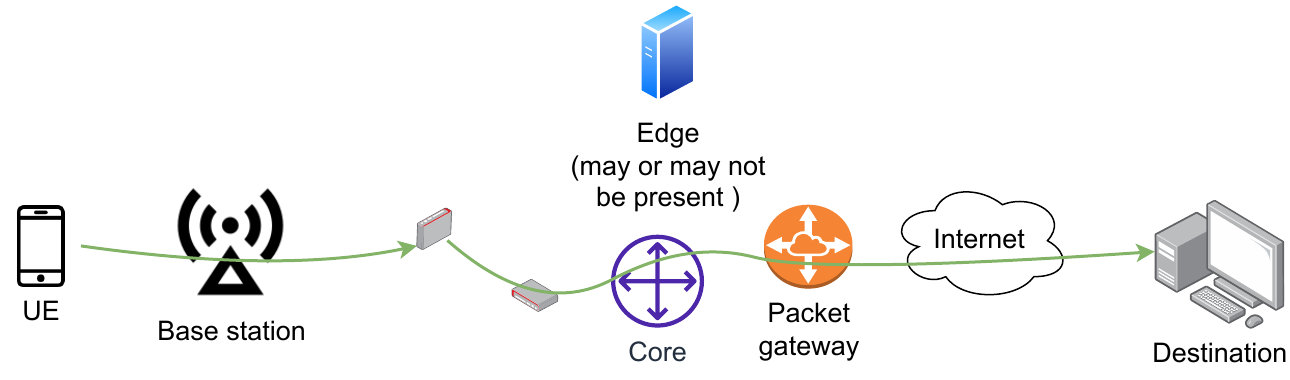}
\caption{5G architecture}
\vspace{-0.2in}
\label{fig:5-g-threat}
\end{figure}

\textbf{Mobile networks:}
In mobile networks, users' traffic must first pass through 
  the radio access network (RAN)
  and the carrier's IP backhaul network, before reaching the Internet.
A typical architecture of a mobile network
  is shown in \autoref{fig:5-g-threat}.
A mobile data network has components connecting
  the user equipment (UE), base station, and 5G-Core (5GC, also known as Evolved Packet Core (EPC) in 4G) 
  to reach the gateway, and then to the IP network.
User equipment (UE) connects to a base station
  through a wireless Radio Access Network (RAN).
The RAN communicates over a radio channel, with
  4G LTE using less than 6\,GHz, but 5G having more than 30\,GHz channels
  in mmWave~\cite{5g-freq}.
Most mobile carriers in the United States
  promise to provide 5G coverage in metropolitan areas,
  but even within metropolitan areas 4G and 5G co-exist
  to ensure backward compatibility.

Edge computing is a new architectural feature in 5G networks,
  where services such as CDNs are placed inside or immediately adjacent to the mobile operator’s network (\autoref{fig:5-g-threat}).
5G suggests that a widespread deployment of mobile edge computing can
  reduce latency and increase throughput.
We may observe performance variation depending on the edge computing's location.

\comment{The following para is rewritten by Plonka to address Akamai legal comments. The prior para is commented out. ---asmrizvi 2024-03-13}
\textbf{Delivery Networks:} To minimize client latency, Content Delivery Networks (CDNs) sometimes place their servers geographically close to the users. Ideally, UE that request for web or streaming services, get their content expeditiously through match-making methods that pair them to nearby servers. 

The path from the base station to the edge computing or
  packet gateway are normally tunnels over an IP network (or they can just be routed normally),
  often it is known as the backhaul network~\cite{5g-archi}.
The base stations can have wired or wireless connections
  to the core and packet gateway~\cite{zhang2019overview}.
Backhaul network contributes to the observed latency
  from the user device.

The relative distance and interaction among
  the UE, backhaul network, IP network,
  and the location of the destination CDN servers
  have an impact on the end-to-end latency.
In this paper, we investigate end-to-end latency 
  when mobile UE interacts with a CDN.

\section{Data Sources And Measurements}
    \label{sec:data-sources}

We use two datasets 
  to characterize end-to-end latency and
  latency inside mobile networks.

\subsection{CDN HTTP Statistics}
    \label{sec:cdn-http}

\begin{table*}
\centering
\begin{tabular}{cccccc}
\textbf{Carrier} & \textbf{Country} & \textbf{\begin{tabular}[c]{@{}c@{}}Observable from \\ server?\end{tabular}} & \textbf{\begin{tabular}[c]{@{}c@{}}Carrier \\ label (bits)\end{tabular}} & \textbf{\begin{tabular}[c]{@{}c@{}}Geo\\ label (bits)\end{tabular}} & \textbf{\begin{tabular}[c]{@{}c@{}}WiFi or\\ mobile label (bits)?\end{tabular}} \\ \hline
1 & U.S.               & Yes                                                                         & 1 to 32                                                                     & 33 to 40                                                               & 41 to 56                                                                          \\
2  & U.S.              & Yes                                                                         & 1 to 24                                                                     & 25 to 32                                                               & 33 to 36                                                                     \\
3  & U.S.              & No for HTTP traffic                                                         & 1 to 32                                                                     & NATed                                                                  & NA \\

4  & Germany        & Yes     & 1 to 32    & 33 to 40   & 41 to 64  \\
5  & Germany        & No     & 1 to 32    & NATed   & NA  \\

6  & Spain        & Yes     & 1 to 32    & Not found   & 33 to 56  \\

7  & India        & Yes     & 1 to 32    & 33 to 48   & 33 to 40  
\end{tabular}

\caption{IPv6 address pattern from server for different US carriers}
\vspace{-0.1in}

\label{tab:bitpattern}
\end{table*}

\begin{table*}
\centering
\begin{tabular}{ccccccc}
\textbf{Country} & \textbf{Carrier}  & \textbf{\begin{tabular}[c]{@{}c@{}}\# of \\client /48s \end{tabular}} & \textbf{\begin{tabular}[c]{@{}c@{}}\# of\\clients\end{tabular}} & \textbf{\begin{tabular}[c]{@{}c@{}}\# of \\serving /24s \\\end{tabular}} & \textbf{\begin{tabular}[c]{@{}c@{}}\# of\\CDN host \\addresses\end{tabular}} & \textbf{\begin{tabular}[c]{@{}c@{}}Duration\end{tabular}} \\ \hline
U.S. & Carrier 1 & 1,830 & 1,412,325 & 769 & 27,018 & 8.4\,h   \\
U.S. & Carrier 2 & 1,327 & 1,416,445 & 639 & 21,584 & 8.4\,h      \\  
Germany & Carrier 4 & 409 & 2,540,339 & 419 & 8,579 & 24\,h     \\
Spain & Carrier 6  & 246 & 620,969 & 211 & 2,840 & 24\,h     \\
India & Carrier 7 & 7,709 & 4,901,684 &  574 & 9,139 & 9\,h \\
\end{tabular}

\caption{CDN dataset in numbers}
\vspace{-0.2in}

\label{tab:dataset-numbers}
\end{table*}

We observe 5G performance from  a global, commercial CDN.
This CDN provides both web and streaming data and hosts DNS.
Our goal is to determine latency and throughput distribution
  from 5G devices.
We observe both client and server-side data between the CDN to 5G devices.
  
\subsubsection{CDN Logs from Server Side}
    \label{sec:cdn-server}
From server side,
  we analyze server logs of sampled HTTP(S) sessions.
The CDN receives millions of HTTP GET requests every second.
The CDN collects a 1\% sample on a specific day, but
  this sampled dataset is large enough with about a billion samples per day.

The CDN samples sessions at the servers.
For each sample we identify the client’s IP prefix, 
  BGP origin AS number, and the server’s IP address.
From the client's origin AS number and IP prefix we identify its provider.
We identify server physical locations from its IP address and CDN internal records.
For each TCP connection,
  the log reports the number of packets,
  information about the round trip time (RTT), bandwidth,
  connection protocols, and congestion.

We analyze data from multiple
  countries to understand global trends.
\autoref{tab:bitpattern} shows the
  carriers from different countries.
We identify the use of Network Address [Port] Translation (NAT) and 
  non-NAT for IPv4 and IPv6 addresses
  from address assignment patterns, 
  as verified with data from devices inside the carriers.
We choose five carriers from four different countries, 
  each with non-NATed IPv6 addresses, to examine (\autoref{tab:dataset-numbers}). 
In the logs for these five carriers, 
  the clients' full IPv6 addresses are visible since they are non-NAT addresses.
From \autoref{tab:dataset-numbers}, we observe over 1\,M unique IPv6 UE
  for each of the carriers.
Each carrier uses 246 to 7,709 /48 IPv6 prefixes (shown by the \# of client /48s column in \autoref{tab:dataset-numbers}).
The number of CDN host addresses 
  with which these clients interact varies, 
  of course, as shown in \autoref{tab:dataset-numbers}.
For instance, Carrier 1 was served from 27,018
  unique IPv4 server addresses. 
  
\textbf{Different RTTs collected by the CDN:} 
From the CDN data, we examine RTTs collected by CDN in two different ways.
First, we get RTTs passively from TCP handshakes where the server kernel
  reports the RTT from SYN-ACK and ACK packets.
Second, TCP reports mid-flow RTTs
  when an ACK arrives and is not discarded.
These RTTs generate the statistics
  from multiple TCP ACKs received by the servers.
TCP handshakes provide a single data point but
  TCP data-ACK RTTs provide multiple observations during a session
  to measure the minimum, maximum, and mean RTT along with the variance within that session.

\subsubsection{CDN Logs from Client Side}
    \label{sec:cdn-client-side}

To complement server-side logs
  and to show the difference between 4G and 5G 
  observed latency and throughput,
  we use real-time logs measured
  from UE to different CDN-hosted services.
CDNs collect these performance logs from user devices
  to evaluate the network condition and to find out 
  the places for improvement.
We use the detailed device and connection information,
  along with the latency data from user devices
  collected by the CDN's real-time 
  user monitoring system.
This dataset reports access network information which
  helps us to distinguish UE using 
  WiFi from those using mobile data networks (\autoref{sec:cellvsnon}).

\subsection{UE-based Measurement}
    \label{sec:direct-ues}

To complement CDN client logs (\autoref{sec:cdn-client-side}) and
  to analyze the stability for a longer duration,
  we measure latency from real UE.
While the CDN collects client logs
  data from real UE,
  they do not contain continuous measurements.

These measurements from UE include latency 
  for transactions with multiple targets in various timeframes.
We use a Samsung Galaxy A52 device
  with 5G capabilities to evaluate latency stability.
We use AT\&T carrier for this measurement.
Unlike the CDN-collected data from UE,
  using our own UE we can collect data 
  for longer duration with our own control.

\section{Methodology: Identifying Mobile Devices and Stability Analysis}

Before we use CDN logs (\autoref{sec:cdn-server})
  to characterize end-to-end latency, throughput, and stability,
  we must understand what the CDN is observing.
A CDN receives traffic
  from many clients,
  so our first goal is to identify mobile UE in the data.
Traffic source IP addresses may identify the originating AS
  as a mobile operator,
  but mobile operators may support a mix of clients using mobile data,
  WiFi, and even wired networks.
We next describe how we use IPv6 address pattern
  to distinguish access network (\autoref{sec:ipv6-to-get-cellular}),
  and how to differentiate 
  4G and 5G (\autoref{sec:diff-4g-5g}).
Finally, we describe how we examine the stability of latency (\autoref{sec:method-stability}).

\subsection{Identifying Mobile UE from IPv6 Addresses}
    \label{sec:ipv6-to-get-cellular}

We use patterns in IPv6 addresses to
  identify a UE's access method (mobile or WiFi)
  and its geographic location.

Mobile providers use
  both IPv4 and IPv6 address space for their clients.
Clients who use IPv4 addresses normally use carrier-grade NAT,
  often mixing clients using many different access network technologies
  into the same IPv4 prefix used by the NAT\@.
However, we find that IPv6 addresses are usually unique for each specific UE.
We therefore use IPv6 addresses 
  so that we can readily distinguish and characterize individual clients' traffic.

Even for IPv6, the CDN sees a mix of NATed and non-NATed addresses.
Some carriers use NAT even for IPv6, 
  in which case we see only translated addresses at CDN servers.
Unfortunately, carriers using IPv6 NAT seem to do NAT at the edge of their network,
  hiding internal structure of the internal UE IPv6 address
  that may offer a clue of the access technology.
Also, the latency to these NATed addresses may not
  represent the actual end-to-end latency to the client if the NAT is also doing split-connection TCP  or using a web proxy.
By contrast, non-NATed addresses 
  show end-to-end latency in CDN server logs and 
  imply that no web proxy is being used.

\comment{this para is true statements, but: WHY are you telling me this?  Is this a method you're going to use to detect NAT?
Or are you just giving some observations?  If it's just observations, maybe swap the last sen from the prior para with the first sen of this para,
so all the non-NAT stuff is in one place.
If you're going to use this to detect, please add a sen saying that.
(Something like ``We show below how we use this trend to identify NATed IPv6 addresses.'' 
---johnh 2024-03-10}
\comment{modified. ---asmrizvi 2024-03-10}
We show below how we discriminate NAT from non-NAT client IPv6 addresses.
A typical NATed address heavily aggregates traffic
  behind an address since many clients behind the NAT use the same address.
However, with a non-NATed address, the traffic is
  significantly lower than the traffic from a NATed address.
We can also observe more individual client IPs
  when there is no NAT.
Also, split-TCP connections result in unrealistic and consistent
  end-to-end latency in the CDN logs,
  as they originate from the same NAT location.
For our analysis, we choose carriers where 
  the query frequency from each source IP is notably 
  lower compared to the query frequency observed with 
  carriers using NATed addresses.

\comment{after reading the last para about NAT, my next question is: ok, so HOW CAN YOU TELL if an address is NATted or not?  you don't say.
If you can't tell, then why does any of that matter? ---johnh 2024-02-23}
\comment{I added the above paragraph. Does that make sense? ----asmrizvi 2024-02-25}
\comment{ NO because you don't actually SAY what you're going to use to distinguish, or how! Please finish the thought. ---johnh 2024-03-10}
\comment{modified the prior para. ---asmrizvi 2024-03-10}

\comment{so after this ``introduction'' you go into 4 subsubsections.
I have no clue what you're going to say or why.
Can you put a statement here about what your plan is for the rest of this section? ---johnh 2024-03-10}
\comment{added. ---asmrizvi 2024-03-10}

Next,
  we show how we use IPv6 address pattern
  to identify carrier, geolocations, and access networks.

\subsubsection{Carrier Labels}

Mobile operators assign UE to fixed subsets of their IPv6 address space.
A prior study also discovered IPv6 address patterns to identify
  client addresses and packet gateways~\cite{zhang2021inferring}. 
In this paper,
  we add to this work by identifying patterns in three non-US carriers.
We also demonstrate how address patterns provide insights 
  into geolocations, WiFi networks, and mobile carrier identities.
  
\comment{If you're building on prior work you ABSOLUTELY MUST CITE that prior work,
 AND YOU ABOSLUTELY should say what you add! 
I guess you now cite them.  suggest s/complement...number of carriers/add to this work by identifying patterns in two non-US carriers./
Then I can't tell: are geolocation, wifi, and cellular vs. wifi YOUR work or theirs? 
PLease be explicit what you add and what they did! 
This is critical to your contribution (and giving them appropriate credit!). 
In addition to saying what we do is new here, for each of the subsections, if it's new, you should say ``We are the first to identify xxx...'' ---johnh 2024-03-10}
\comment{modified and added the last line. ---asmrizvi 2024-03-10}
From \autoref{tab:bitpattern}, we can see that
  among three popular US carriers,
  we find two carriers where the IPv6 addresses of the UE interfaces are directly
  observable from the CDN servers.
The other carrier uses NATed IPv6 address for their HTTP(s) traffic.
Among the two carriers from Germany,
  we observe one with a NATed IPv6 address from the server logs.
For each address, the /24 or /32 IPv6 address prefix identifies the carrier.
Each carrier has different labels (subsets of contiguous bits in the client address), and these
  labels can be the identifier of a carrier.

\subsubsection{Geolocation Labels}

We have identified patterns of geolocation 
  in all four carriers that do not use IPv6 NAT, including two non-U.S. carriers.

We confirm geolocations are consistently used by comparing
  our knowledge of carrier geolocation prefixes
  with CDN server location,
  and from ground-truth locations of specific UE
  from client-side data  (\autoref{sec:cdn-client-side}).
We see the two non-NATing U.S. carriers
  use the middle of their IPv6 addresses for consistent geographic regions.
For example, Carrier-1 uses 8 bits for the geolocation,
  and we consistently see UE in California with one label and
  those in New York with a different label.

\subsubsection{Access Network Technology}
    \label{sec:cellvsnon}

We also find non-NATing carriers use a label
  to identify access network type as mobile or WiFi.
For example, we find a carrier that uses fixed 4 bits label to distinguish 
  mobile and WiFi access network (Carrier-2 in \autoref{tab:bitpattern}).
However, we did not find such characteristics for
  carriers with NATed addresses.
With non-NATed carriers,
  a fixed label is used for either mobile
  or WiFi access network, but not for both mobile and WiFi.
We validate this finding based on the
  measurement from real user devices.
To make sure that we are identifying mobile access network correctly,
  we only identify an IP as a mobile device IP
  when we have ground truth about it,
  measured from real user devices.
To get the ground truth, we utilize
  CDN's measurement from real user devices with
  device information mentioning mobile or WiFi networks as the current access network (\autoref{sec:cdn-client-side}).

From \autoref{tab:bitpattern},
  we can see different patterns for each carrier, 
\autoref{tab:dataset-numbers} shows 
  the non-NATed carriers and
  these carriers' total number of UE
  that we identify from the CDN data.

\subsubsection{Apparent HTTP(S) Proxying} 
We found one carrier that apparently proxies HTTP(S) traffic.
While non-HTTP traffic appears to come from end-device IPv6 addresses,
  HTTP(S) traffic comes from different, NATed addresses,
  and is identifiable by a fixed address pattern.

To confirm this implied proxying is only for HTTP(S), 
  we started HTTP service on three different ports: 80, 443, and 8500.
We found that when the service is open
  at port 80 or port 443,
  Carrier-3 of \autoref{tab:bitpattern} uses a NAT,
  hides the real IPv6 address,
  and sends the requests from the NATed IP address.
From server,
  we can only see the addresses after NAT translation
  for HTTP(S) traffic.
However, when the requests go to a different port (like 8500),
  we observe the unique IPv6 client addresses.

\subsection{Distinguishing 4G and 5G}
    \label{sec:diff-4g-5g}

\begin{table}[]
\centering
\begin{tabular}{c|cc}
\textbf{\begin{tabular}[c]{@{}c@{}}Device/\\ Coverage\end{tabular}} & \textbf{4G area} & \textbf{5G area} \\ \hline
\textbf{4G device}                                                 & 4G               & 4G               \\
\textbf{5G device}                                                 & 4G               & 5G              
\end{tabular}
\caption{Observing 4G and 5G network with respect to device type and network coverage}
\vspace{-0.2in}
\label{tab:4g-5g}
\end{table}

We show how address patterns can tell us about
  the carrier names, geolocations, and sometimes
  network type---mobile or WiFi.
But
  we did not find any evidence in the address pattern
  that can tell us whether the address is
  from 4G or 5G networks.
Often 4G and 5G co-exist in the same physical locations,
  and are supported by the same UE.
Devices can move from one to another
  without changing the address pattern.
Also, some configurations of 5G networks use 
  a software stack composed largely of legacy 4G protocols
  (non-stand-alone mode), since stand-alone
  5G has not been deployed yet widely~\cite{liu20205g}.

The difference in architecture and co-existence of 4G and 5G raises the question, 
  ``can we distinguish 4G and 5G operation?''
We suggest that performance can identify 5G use.
We show possible combinations in \autoref{tab:4g-5g}.
Our first expectation is that a 4G-only device
  can only experience 4G, irrespective of the
  network coverage in that location.
Our second expectation is that a user with a
  5G device may or may not experience 5G capabilities
  depending on the 5G coverage within an area.

Based on this expectation, we use \emph{observed latency} to distinguish
  between 4G and 5G devices.
While latency may vary, if we look at the \emph{minimum} latency for each device,
  we hypothesize that the minimum for a 4G-only device will be higher
  than the minimum for a 5G-enabled device.
We validate that this method works using 8 device models with known capabilities
  (some 4G-only and some 5G-capable)
  in \autoref{sec:result-4g-5g}.

\subsection{Measuring Latency Stability}
    \label{sec:method-stability}

Latency stability means that observed 
  latency is consistent, \textit{i.e.,} determines how much jitter occurs.
Stable latency benefits transport protocols' performance as it often corresponds with the receipt of packets in order (to manage data buffering) and eliminating unnecessary retransmissions (due to incorrectly inferred packet loss), resulting in better user experiences.


Unfortunately, evaluating stability of latency
  from CDN logs is challenging for two reasons.
First, the client IPv6 addresses are not necessarily stable over time, 
  to subsequent sessions, due to dynamic address assignment practices~\cite{rfc4862}.
Second,
  the sampled CDN logs may miss subsequent new flows from the same IPv6 address.

To overcome these two challenges,
  we utilize long-lasting TCP connections and
  measurements from real devices to evaluate stability.
While CDN logs are sampled, we can observe multiple entries in the CDN logs for the same connection
  when there is a long-lasting TCP connection.
We consider a TCP connection long-lasting when
  the connection exists for more than 30\,minutes.
Additionally, we observe stability in the minimum latency.
Since the routing path from the source to the destination
  should usually be stable,
  the minimum latency is expected to be ``often observed''.
We use the term ``stability'' to show 
  how frequently the minimum value appears in
  the observed latency.

For the CDN logs,
  we divide the whole duration of each TCP connection
  into multiple time windows.
We pick the long-lasting TCP connections
  (that are over 30 minutes), and then
  divide the whole duration into windows, each having duration $W$.
Then we find the minimum latency in each time window and 
  calculate the stability within these minimum latencies (more on latency parameters in \autoref{sec:stability-mobile-network}) 
  in different time windows.
We use 10\,minutes for $W$ when we 
  measure stability from CDN logs.
\comment{Prior para with 5s frequency is deleted because we deleted that result.---asmrizvi 2024-03-11}
To complement the CDN log-based latency assessment, 
  we also measure stability for a 5G device, across three weeks.

\section{End-to-end Results: Latency, Throughput, and Stability}
    \label{sec:stability-mobile-network}

In this section, 
  we characterize end-to-end 
  latency, throughput, and stability.
First, we find out the end-to-end latency
  measured from a CDN.
Our key question is: does 5G meet its target of achieving ultra-low latency,
  and high throughput? 
We find that the achieved end-to-end latency 
  can be as low as 6\,ms---not at the target of 2\,ms~\cite{jun2020ultra}, 
  but close.
Mobile clients are also able to achieve throughput exceeding 100\,Mb/s, representing a notable advancement towards delivering high-throughput mobile services.


\subsection{How Low is the Latency?}
    \label{sec:latency}

\begin{figure}
\centering
\includegraphics[width=0.75\linewidth]{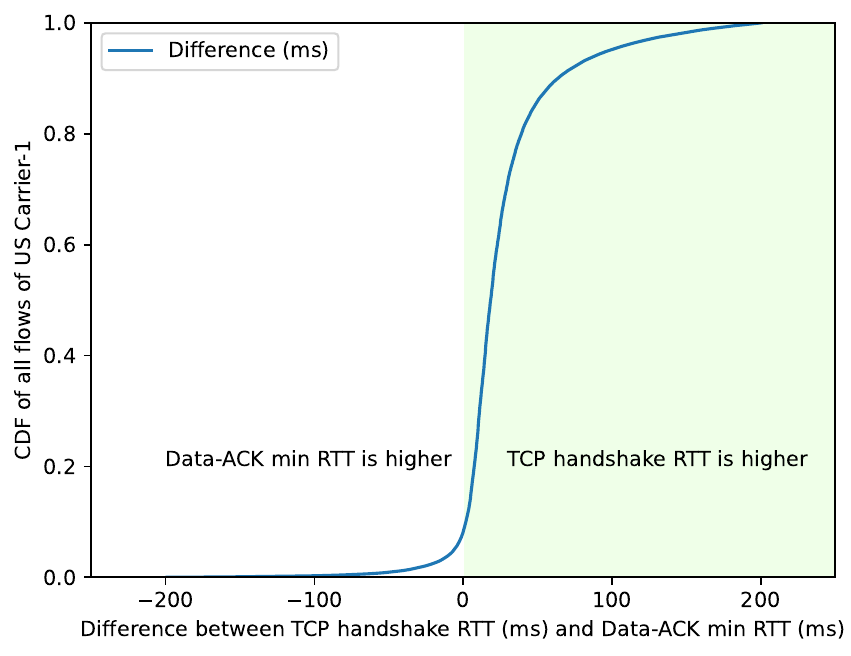}
\caption{Difference between TCP handshake RTT and minimum RTT from data-ACK}
\vspace{-0.2in}
\label{fig:diff-tcp-data-ack}
\end{figure}

\begin{figure*}
	\centering
        \begin{minipage}{.33\linewidth}
		\centering
                \includegraphics[width=6cm]{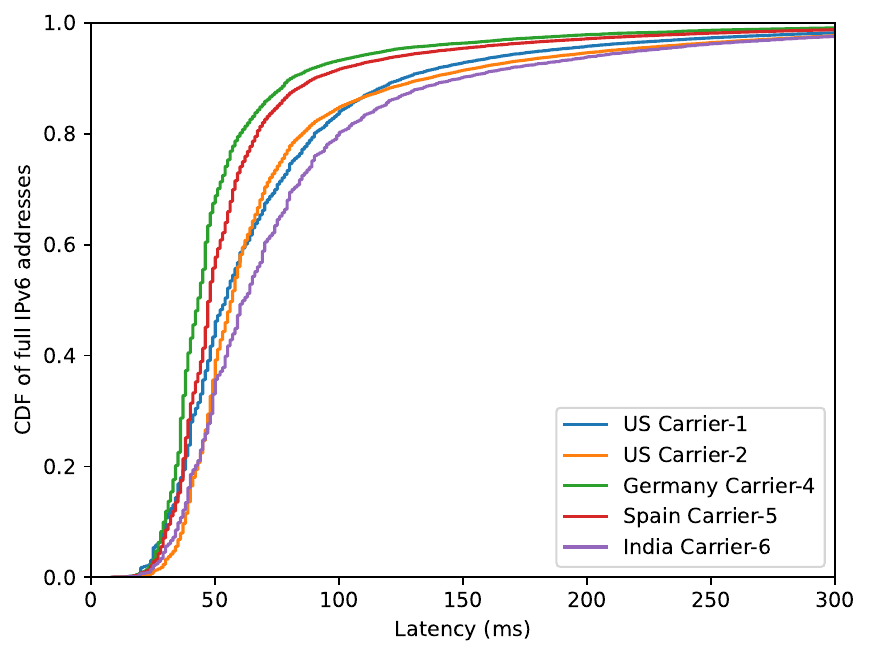}
                \subcaption{CDF of RTT(ms) from TCP handshakes}
                \label{fig:cdf-rtt-tcp} 
	\end{minipage}
	\begin{minipage}{.33\linewidth}
		\centering
                \includegraphics[width=6cm]{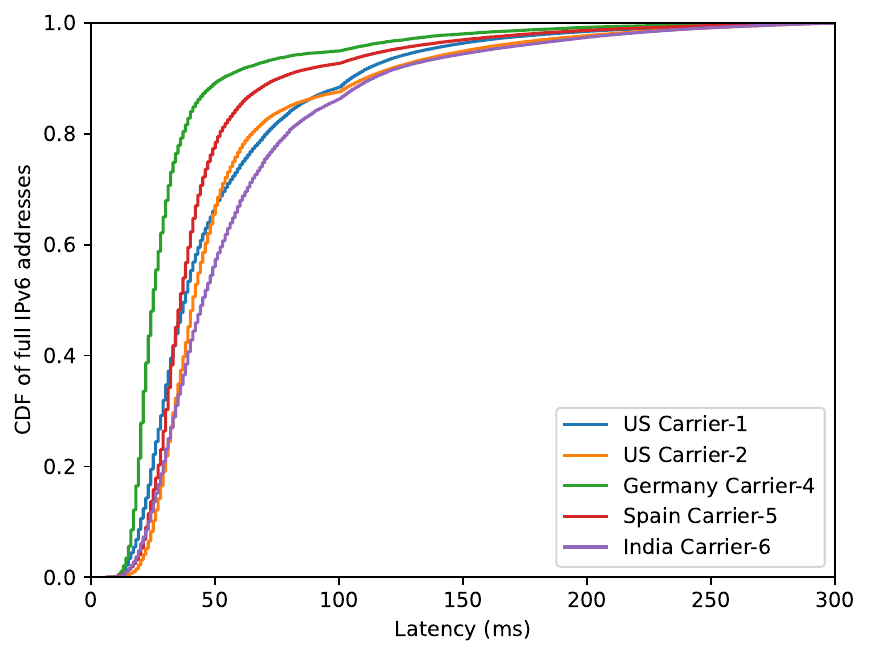}
                \subcaption{CDF of minimum RTT(ms) from ACKs} 
                \label{fig:cdf-rtt-min} 
	\end{minipage}%
        \hfill
	\begin{minipage}{.33\linewidth}
		\centering
                \includegraphics[width=6cm]{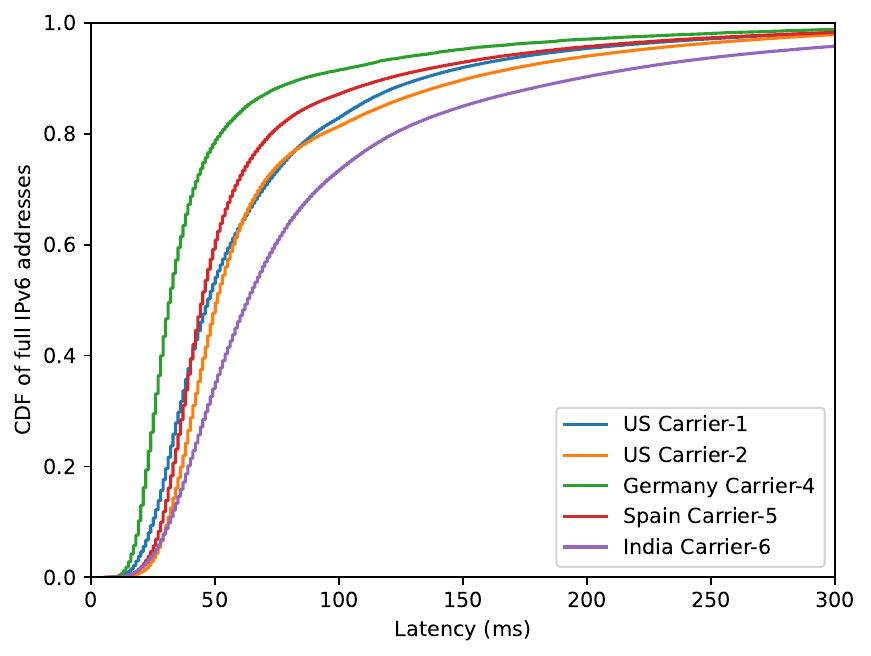}
                \subcaption{CDF of mean RTT(ms) from ACKs} 
                \label{fig:cdf-rtt-mean} 
	\end{minipage}
        \hfill

        \hfill
        \caption{CDF of RTT (ms) in different countries}
        \label{fig:cdn-min}
\end{figure*}

\begin{table*}
\centering
\begin{tabular}{ccccccc}
\textbf{Carrier} & \textbf{Country} & 
\textbf{\begin{tabular}[c]{@{}c@{}}Long-lived\\ TCP conns\end{tabular}} & \textbf{\begin{tabular}[c]{@{}c@{}}Min \\(TCP handshake)\end{tabular}} & \textbf{\begin{tabular}[c]{@{}c@{}}Top 5\% \\(TCP handshake)\end{tabular}} & \textbf{\begin{tabular}[c]{@{}c@{}}Min \\(data-ACK)\end{tabular}} & \textbf{\begin{tabular}[c]{@{}c@{}}Top 5\% \\(data-ACK)\end{tabular}} \\ \hline
Carrier 1 & USA & 4,876 & 9\,ms & 25\,ms & 6\,ms & 17\,ms    \\
Carrier 2  & USA & 3,561 & 8\,ms & 34\,ms & 6\,ms & 22\,ms      \\  
Carrier 4  & Germany & 160 & 12\,ms & 27\,ms & 8\,ms & 15\,ms       \\
Carrier 6  & Spain & 761 & 9\,ms & 28\,ms & 7\,ms & 20\,ms       \\
Carrier 7 & India & 42,516 & 8\,ms &  30\,ms & 6\,ms & 20\,ms  \\
\end{tabular}

\caption{Latency (ms) of the top clients in different countries}
\vspace{-0.2in}
\label{tab:latency-top}
\end{table*}

We first examine latency
  to report the best performance we see today.
We will report two kinds of latency:
  handshake latency (from the connection setup's  initial SYN / SYN-ACK / ACK  exchange),
  and then data-ACK latency extracted during data exchange for TCP.
Each connection provides one estimate of handshake latency
  and many of data-ACK latency,
  so we report CDFs over all connections by a carrier for
  handshake latency, and minimum and mean data-ACK latency.
(We recognize that median is more robust than mean given outliers, 
  the logs contain only the mean within the data-ACKs.)

\textbf{Comparing metrics:}
At first, we start with handshake latency,
  since it is the easiest and most commonly used latency measurement method.
We find the minimum handshake latency is low.
We see the CDF of handshake latency of different carriers
  from different countries in \autoref{fig:cdf-rtt-tcp}.
We find handshake latency can be as low as 9\,ms and 
  the 5{$^{th}$} percentile latency is between 25\,ms to 34\,ms (\autoref{tab:latency-top}).
So, clients that are close to the CDN server with a good 5G coverage
  can expect to observe handshake latency less than 30\,ms.
On the other hand,
  over 50\% of the clients observe more than 40\,ms of TCP handshake latency.

While TCP handshakes are commonly used, 
data-ACK latency measurement is more robust because
  it considers multiple observations over
  the connection lifetime, rather than a single observation at the connection start.

\comment{I don't get ti: you set up mean, but now you do't talk about it, but you go back to ocmparing handshake vs. data-ack?  PLEASE MOVE THIS OR CLARIFY it's about mean. ---johnh 2024-03-10} 
\comment{I am moving it here before the results. Before this section was after the results with mean. ---asmrizvi 2024-03-11}
We show the difference between the RTTs measured from TCP handshakes and 
  data-ACK packets in \autoref{fig:diff-tcp-data-ack}.
We measure the difference between TCP handshake RTT and data-ACK minimum RTT
  for the same flow.
On the green side to the right, 
  the TCP handshake RTT exceeds the minimum RTT for data-ACK packets. 
Conversely, on the no-color side to the left, 
  the minimum RTT from data-ACK packets surpasses the TCP handshake RTT.
Overall, in fewer than 5\% of flows, 
  we note that the TCP handshake RTT is lower than the minimum RTT
  from data-ACK packets.
In 50\% of the flow, the minimum from data-ACK packets and TCP handshake RTT
  have less than 10\,ms of deviation.

With data-ACK latency, we observed 6\,ms as the minimum latency, and 
  the top 5{$^{th}$} percentile latency is between 15\,ms to 22\,ms (\autoref{fig:cdf-rtt-min} and \autoref{tab:latency-top}), when
  the minimum TCP handshake latency is 8\,ms and the 5{$^{th}$} percentile latency is between 25\,ms to 34\,ms.

Finally, we also examine the CDF of mean data-ACK latency.
Because this mean reflects all observations over the flow lifetime,
  it captures variation in latency,
  and when mean is much larger than 5$^{th}$ percentile, it suggests high variance in latency (\autoref{fig:cdf-rtt-mean}).

\textbf{Variation by country:}
In all the countries,
  the median latency is around 50\,ms.
This 50\,ms median is sufficient for most web applications
  but only the top 5\% to 10\% clients would have a better
  experience for latency-sensitive applications.

Among all the countries,
  the German mobile carrier shows a narrower distribution; more clients observe similar latency.
We find 50\% of the German clients
  observe minimum latency of 25\,ms or less (\autoref{fig:cdf-rtt-min}).
The tail latency for India and U.S. carriers is long.
Around 15\% of the U.S. and Indian UE
  observe more than 100\,ms of minimum latency.
The large geographic area of these countries means that
  propagation delay can be large for some UE.
On the other hand, 
  Germany and Spain have a lower tail latency.
Less than 5\% of the UE observe more than 100\,ms of latency.
The Indian mobile carrier has the highest jitter 
  (over 70\% clients show more than 50\,ms of mean in \autoref{fig:cdf-rtt-mean}).
However, in all the cases, similar minimum and 5$^{th}$ 
  percentile latency ensures global 5G deployment and 
  CDN proximity to the mobile users.

\subsection{How Good is Throughput?}
    \label{sec:throughput}

\begin{figure}
\centering
\includegraphics[width=0.75\linewidth]{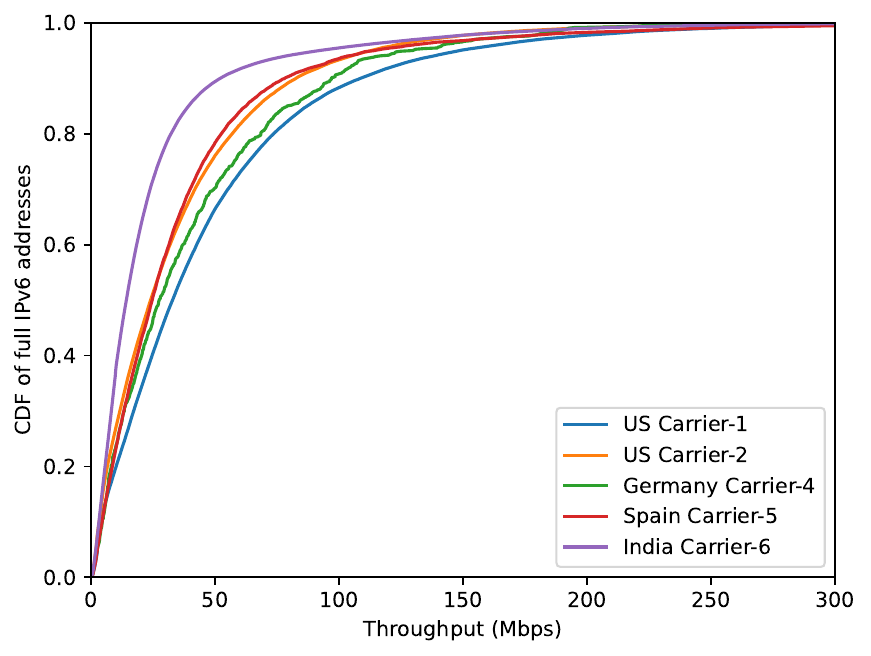}

\caption{CDF of throughput}
\vspace{-0.2in}
\label{fig:through}
\end{figure}


We measure the throughput from the
  transferred bytes and transfer duration,
  assuming uniform transfer speeds.
Although transfer speed may vary over a connection, this method
  estimates actual observed throughput.
We only consider the TCP sessions
  where more than 1\,MB of data is transferred,
  since shorter transfers may underutilize channel capacity due
  to startup overhead (TCP slow start).
We conclude that the observed 5G throughput of 100\,Mb/s 
  is still far under the advertised peak of 20 Gb/s~\cite{5g-performance, pogge2019enabling, an2023adaptive}.

We compute the maximum throughput over all flows for each UE, and 
  show the CDF of this value over all UE.
We show the 
  throughput distribution in \autoref{fig:through}.
The median UE from all carriers sees 40\,Mb/s effective throughput or less.(\autoref{fig:through}).
In the best case of U.S. Carrier 1,
  only 40\% users get more than 50\,Mb/s throughput.
However, some UE see much better performance: 
  the fastest 10\% see 100\,Mb/s or better.

There are several possible reasons UE throughput
  may not exceed 100\,Mb/s:
  insufficient data may not allow the window to open fully,
  either because of small application buffers
  or slow data generation rates by the application,
  or it may represent a bottleneck in either the radio-access or 
  mobile operator's backhaul network.

India has a bigger difference in throughput distribution 
  than other countries (the purple line in \autoref{fig:through}).
Only 25\% of the clients observe more than
  25\,Mb/s throughput for the Indian carrier.
We suspect 5G deployment is still not very mature in India
  or maybe the CDN delivers the content from a distant server.
With the CDN deploying over 9,000 well-distributed server machines in India, 
  we anticipate that the limited maturity of 5G deployment 
  could contribute to low throughput.
\comment{why? ---johnh 2024-02-25}
\comment{added two reasons above. ---asmrizvi 2024-03-01}
\comment{ok.  Given your access to the CDN, can you answer if the second guess is possible?  Does the CDN have servers in India?  Or are you saying maybe the data wasn't staged to that server? 
It would be great to actually get to a reason, not just speculate possible reasons. ---johnh 2024-03-10}
\comment{added the last line. ---asmrizvi 2024-03-11}

To check the impacts of device types over throughput,
  we measure from devices we control, configured to use 4G or 5G only
  with the same mobile provider from
  the same location in Los Angeles County.
We select an iphone 7 as a 4G device and iphone 13 Pro
  as a 5G device, and we put the server within 30 miles
  from the source.
We find up to 30\,Mb/s throughput with iPhone 7, and
  up to 65\,Mb/s throughput with iPhone 13 Pro.
This result shows throughput 
  may vary depending on the device type.
In this controlled experiment,
  we vary the content size, and the server provides enough content 
  so that we can reach the maximum throughput.
Our observed throughput of 65\,Mb/s is within the top 35\% throughput
  that we could observed in \autoref{fig:through}. 
Getting a throughput within the top 35\% of the observed throughput is expected 
  with a 5G-enabled device and within a metropolitan area like Los Angeles.


\subsection{Can We Distinguish 4G and 5G?}
    \label{sec:result-4g-5g}

Do we observe different latency patterns for 4G and 5G devices?
While IPv6 addresses seem to distinguish WiFi from cellular access networks, 
  we do not know how to use them to identify 4G vs.~5G.
Here we use data from user devices collected by
  a measurement system running on user devices that reports the
  device information, access network type, and latency data (\autoref{sec:diff-4g-5g})
  back to the CDN.

\begin{figure}
\centering
\includegraphics[width=0.75\linewidth]{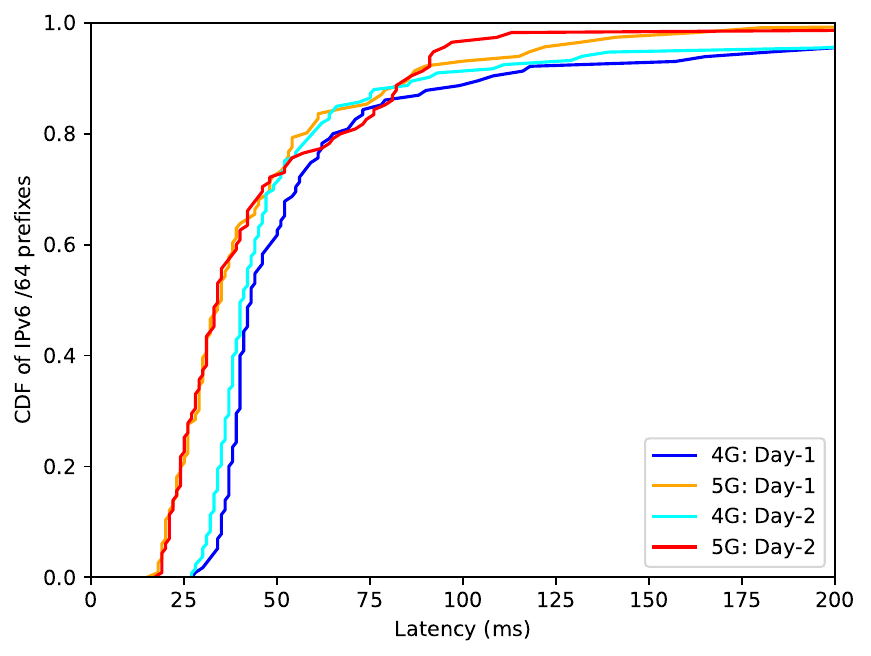}
\caption{Latency observed from 4G and 5G devices}
\vspace{-0.2in}
\label{fig:4g-5g-devices}
\end{figure}

We evaluate the latency from a set of 4G and 5G devices
  located across the U.S.
We select two different sets of user devices---the first one
  is only 4G-enabled, and the second one consists 5G devices 
  (however, existing 5G devices may operate in 4G mode when required).
As 4G devices, we choose Samsung Galaxy S8, Samsung Galaxy S9,
  Samsung Galaxy S10, and Samsung Galaxy A12.
As 5G devices, we choose Samsung Galaxy S21 5G,
  Samsung Galaxy A32 5G, Samsung Galaxy A13 5G, and Samsung Galaxy Note20 5G.
To verify consistency,
  we evaluate latency for two days.
We expect the same set of CDN servers since
  we chose the same web target to compare latency
  from 4G and 5G devices.
We exclude the cases
  when the browser gets the web pages from the cache.

We show that latency \emph{can} distinguish 4G and 5G networks,
  based on latency distributions that we report in \autoref{fig:4g-5g-devices}.
The 4G and 5G devices show a different latency distribution.
We observe the latency data collected from user devices
  to a commercial website hosted by the CDN.
Within a day, we find around 150 unique IPv6 /64 prefixes
  that requested the commercial website.
We observe multiple requests from a single IPv6 /64 prefix.
Multiple requests give us around 400 data points to the 
  target website for a carrier on a particular day.

We find that 20\% of the requests from 5G devices observe less than 25\,ms latency.
On the other hand, no 4G device observes less than 25\,ms
  latency during the two days of our measurements.
5G devices have a wider latency range since
  they may experience both 4G and 5G capabilities.
The tail is similar for both 4G and 5G.
Around 30\% requests
  experience more than 50\,ms of latency,
  which is true for both 4G and 5G.

Our result shows a distinction in the latency distribution
  between 4G and 5G devices.
To confirm that this difference is caused by the
  cellular network technology and not the mobile UE hardware,
  we examined controlled experiences with two devices.
We selected two Samsung Galaxy models with similar hardware
  specifications---Samsung Galaxy A12 and Samsung Galaxy A32 5G.
They both have similar numbers of cores (8 cores each) and CPU clock speeds (2.3\,GHz for Samsung Galaxy A12 and 2.0\,GHz for Samsung Galaxy A32 5G).
\comment{next sen: WHERE do you show this data?  Can you give a NUMBER to support your claim, like ``median latency over all flows is xxx vs. yyy'' ---johnh 2024-03-10}
\comment{I will do it later. Just keeping it for now. ---asmrizvi 2024-03-11}
Comparing the latency distribution between these two models, 
  we find that Samsung Galaxy A32 5G devices
  show better latency compared to the Samsung Galaxy A12.
Since the main difference is 
  phone cellular technology (4G vs. 5G) and not CPU or memory, 
  this comparison suggests that cellular network technology
  can cause latency variation.

While 4G and 5G distributions are different, 
  these distributions overlap, and we do not identify a specific threshold.
If we see latencies below 20\,ms, we identify the device is likely a 5G device
  in a 5G-enabled area,
  however latencies above 50\,ms are common to both 4G and 5G.

\subsection{How Stable is Latency?}
    \label{sec:stability}

Finally, we evaluate the stability of latency
  measured at a CDN.
As outlined in \autoref{sec:method-stability},
  we use long-lasting TCP connections and direct measurements from UE
  to analyze the stability. 

We expect IPv6 addresses to be ephemeral,
  because privacy preserving addresses change frequently, 
  often daily~\cite{rfc4862}.
\comment{i think there are papers that measure ipv6 address stability. you shoudl cite them, in addition to given your own data next. ---johnh 2024-03-10}
\comment{Ok, I will cite them. ---asmrizvi 2024-03-11}
We confirm this result when we look at data where 
  we have UE IPv6 addresses,
  and we see that only 748 of 497,191 IPv6 addresses (only 0.15\%) retain 
  the same IPv6 address after 24 hours.
This almost complete lack of address persistence 
  suggests that UEs frequently change their IP address assignment.
Investigation of latency over time is hampered 
  by dynamic IP addresses because the address 
  has a relatively short client UE association.
Since IP addresses are known to change,
  we examine the stability of latency in long-lived TCP connections,
  since the same TCP connection must go to the same device endpoint.

Next,
  we show the stability of minimum latency at different time windows.

\subsubsection{Evaluating Latency Stability}

\begin{figure}
\centering
\includegraphics[width=0.75\linewidth]{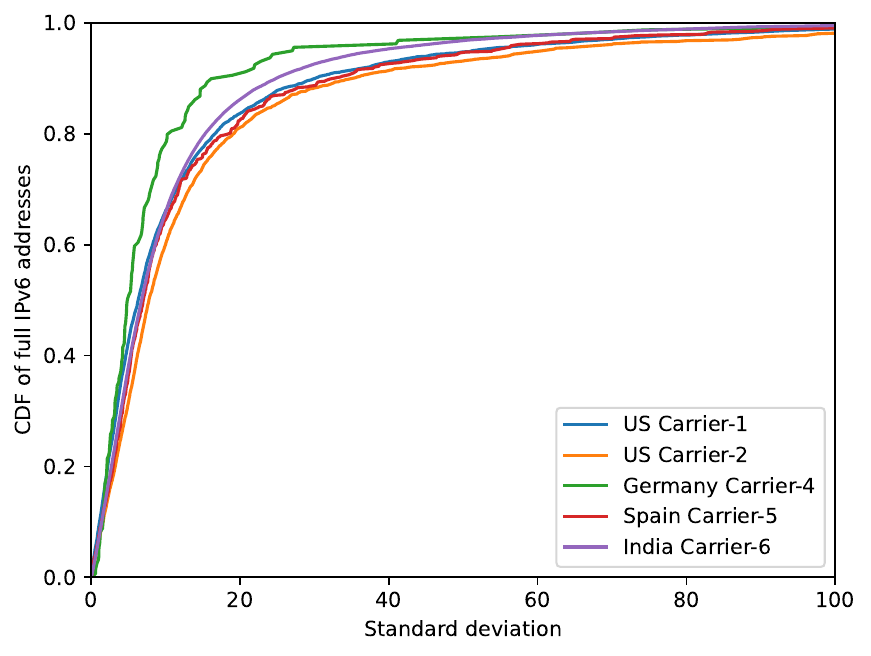}
\caption{Standard deviation among the minimum values}
\label{fig:std-cellular}
\end{figure}

\begin{figure}
\centering
\includegraphics[width=0.75\linewidth]{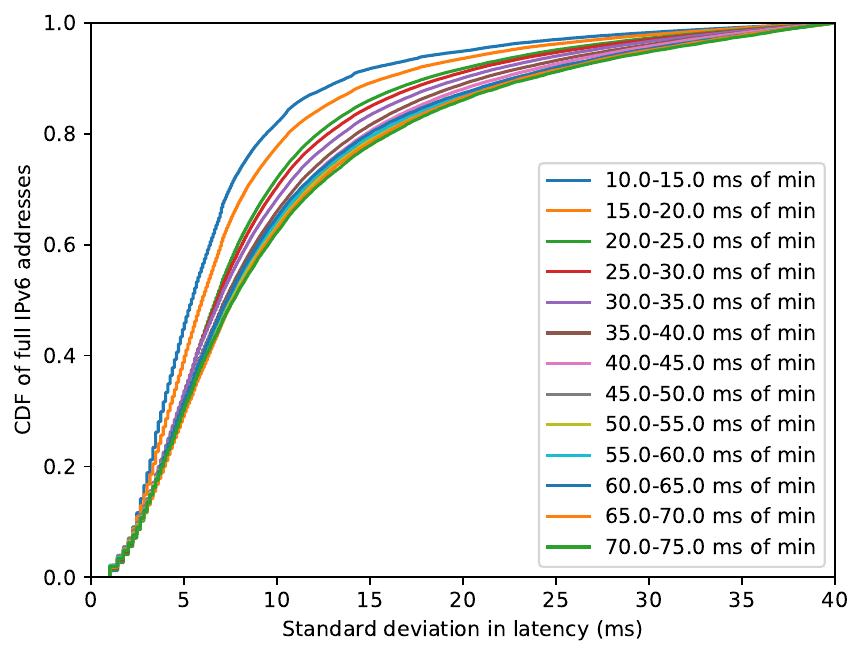}
\caption{Standard deviation among the minimum values}
\vspace{-0.2in}
\label{fig:std-min-cellular}
\end{figure}

By definition, an IP address must remain fixed 
  for a long-lived TCP connection.  
We see a few TCP connections that last 30 minutes or more.

Minimum latency remains stable at different
  time windows for long-lasting TCP connections.
We observe 160 and 761 long-lasting TCP connections for the German and 
 Spanish carriers, respectively; 4,876 and 3,561 for the two U.S. carriers; and 
 42,516 for the Indian carrier (\autoref{tab:latency-top}).
These long-lasting connections are over 30 minutes long.
\autoref{fig:std-cellular} shows the standard deviation
  of minimum latencies in each 10\,minutes window 
  collected from these TCP connections by the CDN.
40\% of these connections show
  less than 5\,ms of standard deviation.
So, long-lasting TCP connections
  show a stable minimum latency for time windows of around 30 minutes.
The global standard deviation in the minimum latency
  measured from the data-ACK packets is low.
In 60\% of the long-lasting TCP connections,
  we observe less than 10\,ms of standard deviation (\autoref{fig:std-min-cellular}).
This is true for all the countries while
  the German carrier shows the highest stability.
We find that 80\% of the German flows observe less than 10\,ms of standard deviation.
Stability in minimum latency
  represents stable end-to-end distance.
There can be many different reasons for unstable latency like
  moving devices, congestion, or maybe poor network coverage.
However, with all these different reasons,
  we observe stable end-to-end minimum latency.

We observe low baseline latency correlates with stability.
\autoref{fig:std-min-cellular} shows how
  minimum latency and standard deviation are related to each other.
We calculate the minimum latency and standard deviation
  among the round trip times (RTTs) measured
  from the data-ACK packets within a TCP connection
  using the ACK packets.
We find that when the minimum latency is low,
  the standard deviation is also low.
When the minimum latency is 10-15\,ms, the
  standard deviation is around 5\,ms for 50\% of the
  IPv6 addresses.
However, the standard deviation is around
  8\,ms for 50\% of the addresses,
  when the minimum latency is 70-75\,ms.
The lines gradually shift to the right (more standard deviation)
  as the minimum latency shifts from 10\,ms to 70\,ms as we can see from \autoref{fig:std-min-cellular}.

\subsubsection{Stability over Three Weeks}
    \label{sec:stability-longer}
\begin{figure}
\centering
\includegraphics[width=0.75\linewidth]{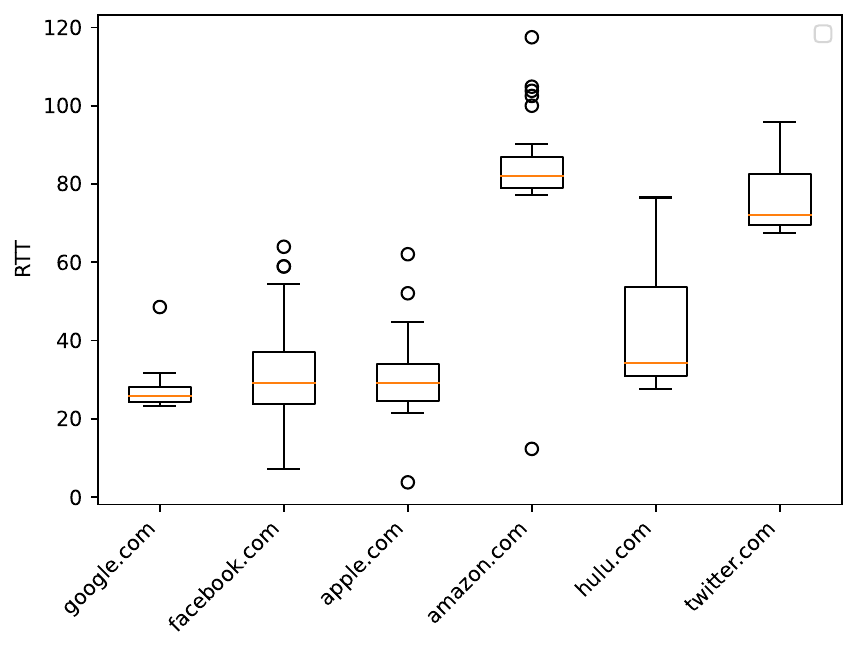}
\caption{Latency from one UE over three weeks.}
\vspace{-0.2in}
\label{fig:panorama}

\end{figure}

We next look at the stability of one device for three weeks.
This data complements our prior examination of 
  thousands of devices for tens of minutes.

To check the stability for an even longer period,
  we measure the latency from a single piece of UE (\autoref{sec:direct-ues}) for over three weeks.
We select different popular top-level popular webpages as our targets,
  and ping these targets everyday from a specific location.
Since we use the domain names as our targets,
  the target IPs may contain both IPv4 and IPv6 addresses.
Then we measure the lowest latency
  within the day to see whether this minimum latency remains stable
  for three weeks.

\autoref{fig:panorama} shows latencies from the same UE to top six websites,
  measured up to 30 times per day for three weeks.
For each site we consider the minimum RTT observed each day,
  then summarize those 21 days with a boxplot showing median and quartiles (the $P_{25}$ and $P_{75}$ percentiles)
  and whiskers showing the largest and smallest values within $1.5\times\V{IQR}$ beyond the $P_{25}$ and $P_{75}$ values, where IQR is $P_{75}-P_{25}$.

Overall, we see latencies vary quite a bit for each site,
  and even more between sites.
While Google, Facebook, and Apple are all consistently around 20 ms, 
  Amazon, Hulu, and Twitter are 2–4$\times$ that value, 
  suggesting some sites are located at or near mobile provider connection points, while others are remote and accessed over longer paths.
\autoref{fig:panorama} shows different range of latencies
  for different targets.
Pinging to google.com gives us the lowest and most
  stable latency from a specific location.

\section{Conclusion}

In this paper, we present a unique
  evaluation of mobile
  latency, throughput, and stability.
We utilize a globally distributed CDN's logs
  and direct measurements from UE to characterize end-to-end latency.
We demonstrate how IPv6 address patterns can help us
  to identify UE with mobile access network.
Then, upon isolating mobile traffic,
  we analyze mobile latency, throughput, and stability from a globally distributed CDN.
We study mobile carriers 
  in four countries and three continents.
We show end-to-end mobile latency can be as low as 6\,ms,
  and exceeding 100\,Mb/s of throughput is not rare
  from a CDN.
We also show minimum mobile latency remains
  fairly stable when the baseline latency is low.
Our measurements and analysis suggest 
  many mobile users are still far from the performance 
  one might expect of this 5G era. 
Ongoing use of our carrier-independent 
  methods may tell if and when this has improved.

\appendices  

\section{Ethical Considerations}
	\label{sec:ethics}

Our work poses no ethical concerns to the best of our knowledge.
Our work contributes to the community by offering insights 
  into the performance of 5G networks reaching a globally distributed CDN.
It poses no risks to individuals or organizations.
We preserve the anonymity of the operators' names since 
  our goal was not to compare cellular networks or scrutinize the CDN provider.
Some of our measurements reexamine data from CDN traffic,
  but we do not access user identities and report only aggregate information.
Our measurements from specific UE are carried out by ourselves,
  with devices we selected, and our consent.

\bibliographystyle{plain}
\bibliography{ref}

\end{document}